\documentclass{article}
\usepackage[T1]{fontenc}
\usepackage[utf8]{inputenc}
\usepackage{ismir} % Remove the "submission" option for camera-ready version
\usepackage{amsmath,cite,url}
\usepackage{graphicx}
\usepackage{color}
\usepackage{multirow}
\usepackage{booktabs}
\usepackage{amssymb}

\title{MEX: The Musical Expectation Benchmark Suite}
\multauthor
  {Mathias Rose Bjare$^1$ \hspace{1cm} Augustin Bouquillard$^{2,3}$ \hspace{1cm} Giorgia Cantisani$^4$}
  {{\bf Seung-Goo Kim$^3$ \hspace{1cm} Stefan Lattner$^5$ \hspace{1cm} Gerhard Widmer$^1$}\\
  $^1$ Johannes Kepler University, Linz, Austria \quad $^2$ École polytechnique, Palaiseau, France \quad
  \\
  $^3$ Max Planck Institute for Empirical Aesthetics, Frankfurt am Main, Germany\\
  $^4$ STMS, CNRS, IRCAM, Sorbonne Université, Paris, France\\
  $^5$ Sony Computer Science Laboratories (CSL), Paris, France\\
  {\tt mathias.bjare@jku.at}
}

\def\authorname{M. R. Bjare, A. Bouquillard, G. Cantisani, S.-G. Kim, S. Lattner, and G. Widmer}

\usepackage[bookmarks=false,pdfauthor={\authorname},pdfsubject={\pdfsubject},hidelinks]{hyperref}
\usepackage{cleveref}
\begin{document}

\maketitle

\begin{abstract}
We propose the Musical Expectation (MEX) benchmark suite for perceptually validating computational models of musical expectation, \textit{i.e.,} models that estimate music information dynamics (MID) measures of perceived musical expectations, such as surprisal or uncertainty from symbolic music or musical audio.
MEX collects and extends tasks for evaluating such models previously proposed in music information retrieval and cognitive sciences using standard datasets, and evaluates whether:
(1) proposed MID measures decrease with repetition due to in-context learning of the computational model,
(2) peaks in such measures align with boundaries of music segments,
(3) pitch surprisal and uncertainty estimated from audio are consistent with those modeled in event-based symbolic music,
(4) proposed MID measures can predict electrophysiological responses to music.
We stress that this benchmark suite does not provide an exhaustive computational characterization of music expectation.
Rather, we envision this as an active, interdisciplinary initiative, with benchmarks evolving as our understanding of the underlying perceptual processes deepens. 
We provide open-access code\footnote{\url{https://github.com/musicperceptmodeling/mex}} to estimate surprisal and uncertainty 
from audio using baseline models, run the tasks on standard datasets, and include utilities for running the tasks with users' own data.
\end{abstract}
%TC:ignore
\section{Introduction}\label{sec:introduction}
% Definition (clarification of the term "")
Computational models of music expectation aim to formalize how listeners form predictions about music continuations based on prior knowledge internalized over the course of their lives, as well as on immediate contextual information \cite{huron2006sweet, idyom}. This capacity for anticipation is thought to be central to the musical experience, underlying the sense of tension and release, memory and emotional responses, as well as reward-related pleasure \cite{huron2006sweet, koelsch2014brain, gold2019predictability, cheung2019uncertainty}. 
Early accounts of musical expectations relied on music-theoretic rules explicitly encoding Western tonality principles (e.g., \cite{feld1984generative, narmour1990analysis, schellenberg1997simplifying, temperley2008probabilistic}).
Yet, Tillmann et \textit{al.} \cite{tillmann2000implicit} showed that knowledge of tonal structure could be implicitly learned by a connectionist model through neural self-organization from mere exposure to tone sequences.
This shift towards implicit learning naturally motivated information-theoretic approaches, originally proposed by Meyer \cite{meyer1957meaning}, in which expectations are formalized as music information dynamics (MID) derived from a statistical model's distribution over possible continuations.
Examples include surprisal, the negative log-likelihood of the continuation, and entropy, which quantifies prediction uncertainty given a musical context.

Markovian models such as IDyOM \cite{idyom} operationalize this framework by providing a principled way to learn statistical regularities from large corpora of symbolic music, helping establish these quantities as effective proxies for human expectation that can predict neural and behavioral responses to music \cite{steinbeis2006role,kim2011effect,hansen2014predictive, moldwin2017statistical,cheung2019uncertainty,gold2019predictability, di2020cortical, bianco2020, cantisani2024neural}. IDyOM, however, can only process univariate sequences, such as melodies or chord progressions \cite{pearce2012auditory}.
\\More recent deep learning approaches have shown promise in estimating note-pitch surprisal that can predict neural responses to melodies (e.g., MelodyRNN \cite{sankaran2024encoding} and the Music Transformer \cite{kern2022cortical}), yet their advantage over more interpretable \textit{n}-grams models like IDyOM remains unclear.
Their main potential rather lies in the capacity to capture the high-dimensional dependencies inherent to polyphonic music (e.g., PolyRNN \cite{robert2024polyrnn}), but their ability to predict human expectations remains largely untested.
While the modeling described above relies on symbolic music, recent work has shown that it is possible to estimate musical expectations directly from sound, hence capturing all complex dimensions of polyphonic music at once, including acoustic aspects, offering a more ecologically valid account of the underlying perceptual mechanism \cite{Bjare2024audioic, Bjare2025diffic, Bjare2025perceptually}.
\\Despite this progress, existing studies are fragmented across datasets, tasks, and evaluation criteria, hindering direct comparisons.
In research on statistical modeling of musical sequences, proposed models are often evaluated solely on predictive performance on hold-out music data.
Although useful for understanding information-theoretic limits on perceived musical expectation, such evaluations alone do not tell how well model-derived expectations align with human expectations. On the other hand, cognitive studies on the neural and behavioral correlates of musical expectations often rely on a single model trained on a specific dataset, without considering alternatives that might better model the music data, potentially leading to erroneous inferences when MID measures are derived from less-calibrated or simplified models.
\\To facilitate standardized model evaluations, benchmarking suites have accelerated progress and reproducible research in other applications by standardizing tasks, data, and preprocessing \cite{rocca2021putting,huang2026addressing}.
In natural language processing,  \cite{lighteval} provides standardized benchmarks with shared tasks, evaluation protocols, and datasets that enable rigorous comparisons across models, with an interface accessible to both linguists and AI generalists.
In music information retrieval (MIR), MIREX has established standardized evaluation protocols for tasks including melody extraction, beat tracking, and source separation, enabling systematic comparisons of algorithms over time \cite{mirex}.
Yet, validating musical expectation models requires expertise in cognitive theory and empirics, which may explain why such benchmarks have not been developed in the MIR field.

We propose a benchmark suite for computational models of musical expectation in audio — the Musical EXpectation (MEX) Benchmark Suite — providing a common ground for facilitating interdisciplinary work. MEX is designed to be (i) \textit{perceptually-grounded}, including validation on human data, (ii) \textit{reproducible}, through publicly available code and data, (iii) \textit{modular}, in that it can expand as our knowledge of the underlying cognitive mechanism alongside its computational modeling progresses.
\\The initial release includes extended versions of previously published tasks and evaluate the degree to which: (RE) repetition reduces MID measures \cite{Bjare2024audioic}, (SB) music segment boundaries align with peaks in MID measures \cite{Bjare2025diffic}, (SY) pitch surprisal and entropy modelled in audio are consistent with those modelled in event based symbolic music  \cite{Bjare2025diffic, Bjare2025perceptually}, (NE) MID measures are neurally encoded in electrophysiological responses to music  \cite{Bjare2024audioic, Bjare2025perceptually}.
We release MEX as a \texttt{Python} + \texttt{Git} package, with the hope of facilitating discussion and continuous development. 

\section{Related Work}
Computational models of musical expectation have progressed across domains of increasing complexity in terms of statistical modeling, from monophonic \cite{idyom_conklin,idyom,BjareLW22}, to polyphonic \cite{bjare2024controlling, robert2024polyrnn} models working in the symbolic domain, and, recently, from sound feature \cite{dubnov2007audio, Skerritt-DavisE18, skerritt2019model} to waveform models in the audio domain \cite{Bjare2024audioic, Bjare2025diffic, Bjare2025perceptually}. In such works, the surprisal and uncertainty aspects of musical expectation are usually proxied by information content (IC) and entropy measures derived from autoregressive sequence models that predict music continuations in either the symbolic or the audio domain. While other MID measures have been proposed (e.g., predictive information-based measures \cite{abdallah2009information}), surprisal and uncertainty are the ones that have been most extensively perceptually validated.

Only a few studies, however, have conducted a systematic comparison across models with respect to perceptual validity. 
For instance, in \cite{idyom, harrison2020ppm}, variants of the same \textit{n}-gram model underlying IDyOM are compared with respect to how well the estimated IC predicts behavioral data across a variety of tasks.
%, which, however, are not publicly available.
Bjare \textit{et al.} \cite{BjareLW22} compare aspects of in- and out-of-context learning for different model classes in terms of the dynamics and adaptation of the estimated IC.
In \cite{morgan2019statistical}, Temperley's rule-based model \cite{temperley2008probabilistic} is compared with IDyOM using a melodic cloze test, a behavioral task for quantitatively measuring note cloze probabilities \cite{fogel2015studying}.
% that is  pusblished.
In \cite{robert2024polyrnn}, the authors compare their proposed polyphonic symbolic model to that of \cite{performance-rnn-2017} in terms of predicting brain responses to polyphonic piano, and to IDyOM for responses to monophonic music.
%, but the experiments are not public available.
% However, IC alone is not a good way to compare anticipatory performance, because IC depends on the underlying data representation (e.g., piano roll or token-based encodings). 
In \cite{Bjare2025diffic, Bjare2025perceptually}, different waveform models are compared in terms of how well the estimated surprisal aligns with that estimated by a symbolic model (i.e., SY task), and how well it can predict neural responses (i.e., NE task).
\\However, the reproducibility and generalization of model comparisons in these studies are constrained by the use of tasks or datasets that are often highly specific or non-public.
In this work, we address this limitation by collecting and standardizing diverse musical expectation data and tasks within a unified, publicly accessible benchmark suite.

\section{Benchmarks}
\subsection{Benchmark Class}
\label{sec:infometandbaselines}
Each benchmark consists of a task and a dataset and can be executed given a one-dimensional, real-valued time series of constant frame rate derived from a music dataset.
The task specifies how the time series, paired with task-specific metadata (e.g., segmentations, neural responses), are processed and aggregated to a task-specific score.
In \Cref{sec:results}, all tasks are run in a zero-shot setting, meaning that the time series is predicted from audio alone and that no task-specific fine-tuning is performed a priori.

\subsection{MID Measures and Baselines}
\label{sec:mmidsandbaselines}
Although any MID measure can be benchmarked, we report baseline results only for IC and entropy estimated by the openly available audio-based models from Bjare \textit{et al.}, namely GIVT \cite{Bjare2024audioic} and DIF \cite{Bjare2025perceptually}.
Both are autoregressive Transformer-based models \cite{vaswani2017attention} that estimate next-frame densities in sequences of autoencoder representations based on Music2Latent \cite{pasini2024music2latent} at a frame rate of 11Hz.
As such, each model $f_\theta$ calculates IC of a frame $x_t$ corresponding to an instantaneous surprise at time $t$ by $
  \text{IC}\left(x_t \vert x_{<t}\right) = -\log p_{f_\theta}\left(x_t \vert x_{<t}\right)
$.
More generally, IC can be calculated for frame spans $x_{t:t+k}$ via the chain rule, corresponding to the surprise of the musical surface within that time span, by
$
  \text{IC}(x_{t:t+k} \vert x_{<t}) = \sum_{t'=t}^{t+k} \text{IC}(x_{t'} \vert x_{<t'}) 
$.
% Likewise, the instantaneous and timespan 
Likewise, the timespan 
\textit{(differential) entropy} corresponding to the uncertainty of the prediction can be calculated as an expectation of the IC:
$
% \begin{align*}
  % \mathcal{H}\left(X_t \vert x_{<t}\right)
  %   &= \mathbb{E}_{X_t}\left[-\log p_{f_\theta}\left(X_{t} \vert x_{<t}\right)\right] \\
  % \mathcal{H}(X_{t:t+k} \vert x_{<t})
  %   &= \sum_{t'=t}^{t+k} \mathcal{H}(X_{t'} \vert X_{t:t'-1},x_{<t'}).
% \mathcal{H}(X_{\!{\scriptscriptstyle t:t+k}}\! \mid\! x_{\!{\scriptscriptstyle <t}})
%  =-\! \sum_{\!{\scriptscriptstyle t'=t}}^{\!{\scriptscriptstyle t+k}}\mathbb{E}_{\!{\scriptscriptstyle X_{t:t'}\mid x_{<t}}}
% % \left[
% \log p_{\!{\scriptscriptstyle f_\theta}}(X_{\!{\scriptscriptstyle t'}}\! \mid \! X_{\!{\scriptscriptstyle {t:t'-1}}},x_{\!{\scriptscriptstyle <t}}).
\mathcal{H}(X_{t:t+k} \mid x_{<t})
 =- \sum_{t'=t}^{t+k}\mathbb{E}_{X_{t:t'}\mid x_{<t}}
\log p_{f_\theta}(X_{t'} \mid X_{t:t'-1},x_{<t}).
% \right]
% &=
% \sum_{t'=t}^{t+k}
% \mathbb{E}_{\substack{X_{t'}\mid H_{t'}}}
% \left[
% -\log p_{f_\theta}(X_{t'}\mid H_{t'})
% \right].
% \end{align*}
$
% Hereafter, we will refer to the differential entropy simply as entropy.
As a hybrid of uncertainty and surprisal, the \textit{pathwise conditional entropy} is given by \cite{e28020195}
% $
% \mathcal{H}_{\!{\scriptscriptstyle p}}(X_{\!{\scriptscriptstyle t:t+k}}\! \mid\! x_{\!{\scriptscriptstyle <t}})
%  =-\! \sum_{\!{\scriptscriptstyle t'=t}}^{\!{\scriptscriptstyle t+k}}\mathbb{E}_{\!{\scriptscriptstyle X_{t'}\mid x_{<t'}}}
% % \left[
% \log p_{\!{\scriptscriptstyle f_\theta}}(X_{\!{\scriptscriptstyle t'}}\! \mid \! x_{\!{\scriptscriptstyle <t'}}),
% $
$
\mathcal{H}_{p}(X_{t:t+k} \mid x_{<t})
 =- \sum_{t'=t}^{t+k}\mathbb{E}_{X_{t'}\mid x_{<t'}}
\log p_{f_\theta}(X_{t'} \mid x_{<t'}),
$
and describes the uncertainty in expectation while being exposed to a fixed context.
In practice, an unbiased estimate of the differential entropy can be calculated using Monte Carlo samples, \textit{i.e.,} by sampling from the underlying generative model and averaging the samples' IC.
In our experiments, we use pathwise conditional entropy, since it avoids error accumulation otherwise occurring in Monte Carlo sampling of frame span entropies \cite{DBLP:journals/corr/RanzatoCAZ15}.

\subsection{Data}
\label{sec:data}
This section summarizes the datasets selected to standardize the benchmarks.
Each contains audio alongside task-specific metadata (e.g., human annotations or neural responses) summarized in \Cref{tab:tasks_datasets}.
Any other audio dataset in which audio is paired with metadata following the same structure could be added in the future.

For the IDyOM consistency task (SY), we extended the synthetic dataset (SYN) from \cite{Bjare2025diffic} to include 200 Irish folk tunes from \textit{The Session} dataset \cite{folkrnnsession}.
Each tune is synthesized at constant velocity with five SoundFont-based instruments---\textit{Pad 1 (new age)}, \textit{Synth Voice}, \textit{Acoustic Guitar (nylon)}, \textit{Acoustic Grand Piano}, and \textit{Trumpet}---yielding 2000 examples in total. For each melody, we also compute IDyOM IC and entropy (see \cite{BjareLW22} for details).
% We publish SYN on \url{https://huggingface.co/datasets/musicperceptmodeling/mex}.
For the neural encoding task (NE), we use the dataset from \cite{cantisani2023investigating} (VOC), which contains electroencephalography responses (EEG) to 18 vocal melodies selected from NHSS recordings \cite{sharma2021nhss}.
For the segment-related tasks, such as repetition (RE) and segment-boundary (SB) detection, we use the SALAMI dataset (SAL) \cite{DBLP:conf/ismir/SmithBFRD11}, which contains 1,310 multi-source audio recordings of different genres annotated by one or two human annotators.
These annotations are hierarchical and include the levels ``\textit{Lowercase}'', ``\textit{Uppercase}'', and ``\textit{Function}'', representing local structure, global structure, and semantically labeled global structure, respectively.
We publish audio and metadata of SYN and NE and metadata of SAL on \url{https://huggingface.co/datasets/musicperceptmodeling/mex}.

\subsection{Tasks}
\begin{table}
\centering
\footnotesize
\begin{tabular}{lll}
\toprule
Task & Metadata Type & Audio Dataset \\
\midrule
RE & Segment annotations & SAL \\
SB & Segment annotations & SAL \\
SY & IDyOM IC and Entropy & SYN \\
NE & EEG & VOC \\
\bottomrule
\end{tabular}
\caption{Tasks, metadata types, and audio datasets.}
\label{tab:tasks_datasets}
\end{table}
\label{sec:tasks}
\subsubsection{Reduction in MID Measures with Repetition (RE)}
Among human-produced signals, music strongly emphasizes repetitions at all time-scales: from rhythmic patterns spanning seconds, to larger sections within a musical piece, and even to repeated listening of the same piece\cite{margulis2013repeat}. Behavioral data show that repetitions not only play a role in our aesthetic appreciation of music \cite{margulis2013aesthetic}, but, most interestingly, can also make a random tone sequence sound like music \cite{simchy2018sound}.
From a statistical learning perspective, repetitions have been shown to reduce uncertainty and refine predictions \cite{pearce2010unsupervised,hansen2014predictive,Bjare2024audioic}.
Hence, a computational model of music expectation is expected to properly internalize repetitions that naturally unfold within a given musical piece.
% How repetitions influence MID measures estimated by a computational mdoel of music expectations is important because listeners formulate predictions based on long-term knowledge but also short-term contextual information, with repeated patterns (e.g., phrases) found to modulate neural responses during musical form processing \cite{pearce2010unsupervised,hansen2014predictive,neuhaus2009similarity}.
\\To quantify the effect of repetitions on MID measures, the RE task follows the approach of \cite{Bjare2024audioic}, and tests to what extent repetitions in music reduce MID measures as a result of in-context adaptation (learning) using SAL.
Here, a repetition is defined as a pair of segment indices in which the second segment repeats the label of the most recent earlier segment with that same label.
For example, a piece with the musical form \textit{A,A,B,A} contains two repetitions, namely (1,2) and (2,4), both associated with the \textit{A} segments.
We then quantify the decrease in MID measures for a repetition as the difference between the average frame-wise values of the two corresponding segments.
For IC and entropy as defined in \Cref{sec:mmidsandbaselines}, this corresponds to time-normalized surprisal and entropy of the segments.
%We collect these differences across all segment types and pieces, compute their mean ($\mu$), and perform a one-tailed \textit{t}-test of the null hypothesis $H_0: \mu \ge 0$ against the alternative $H_1: \mu < 0$.
We collect these differences across all segment types and pieces, compute their mean $\mu$, and test whether it is negative (one-tailed \textit{t}-test).
Because MID measures may lie on different scales across models and measures---e.g., due to differences in autoencoder representations, tokenization schemes, or the unbounded nature of IC---we standardize them beforehand.

Using this repetition definition, model in-context learning should reduce the MID measure on contiguous repetitions like (1,2), and noncontiguous repetitions like (2,4).
This generally requires sensitivity to musical form.
For example, the form \textit{A,A,B,A} is common in pop and jazz music, and familiarity with such styles should help a model anticipate the return of segment A at index 4 after observing the music up to segment B at index 3.
% In addition to considering all repetitions (ALL), we consider a simpler setting that includes only contiguous repetitions (CO). For example, (1,2) is included in (CO), whereas (2,4) is not.
% Finally, we consider (FI), which restricts (CO) to the first occurrence of a contiguous repetition, e.g., only (1,2) in \textit{A,A,A}.
In addition to considering all repetitions (ALL), we consider (FI), which only considers the first occurrence of a repetition.
For example, (1,2) is included in (FI), whereas (2,4) is not.
We include this scenario to investigate cases where the MID decreases on the first repetition but does not necessarily decrease monotonically with subsequent repetitions.
  
\subsubsection{Segment Boundary Alignment (SB)}
% Segment boundaries are a natural test case because listeners rely on changes in the musical surface to perceive musical boundaries, while phrase closure elicits robust boundary-related neural responses; accordingly, both salient boundary changes and violations of expected continuation or closure are plausible sites of elevated expectation-related measures \cite{knosche2005perception, nan2009non, nan2006perception, bruderer2006perception, nan2009non}.
Perception of segment boundaries has been linked to acoustic changes, like sudden timbre,
register, or dynamic changes, that listeners use as cues to infer structural divisions or segmentation \cite{deliege1987grouping,bruderer2006perception}. Moreover, phrase closure has been shown to elicit robust boundary-related neural responses \cite{knosche2005perception, nan2006perception, nan2009non}. Accordingly, both salient surface changes at segment boundaries and violations of expected continuation or closure are plausible sites of increased expectation-related responses.
Hence, the SB task tests whether large changes in the proposed MID measures extracted from audio coincide with segment boundaries, following the approach previously proposed in \cite{Bjare2025diffic}.
We conduct an experiment in which we perform novelty-based segmentation \cite{muller2015noveltysegmentation} by predicting SAL Lowercase, Uppercase, and Function segment boundaries using the most significant peaks extracted from MID measures considered as novelty functions.
We report performance in terms of precision and recall for predictions that are accurate within $\pm1.0$ seconds of the annotations.
As music segmentations can be hierarchical, with potentially larger changes occurring at higher rather than lower levels, and since MID measures might exhibit differences in smoothness, scale, etc., we find $F_1$-optimal parameters for the peak-picking algorithm at each hierarchy level and MID measure using 5-fold cross-validation.
For this work, we use novelty functions constructed by smoothing MID measure novelty functions with Gaussian kernels with standard deviation $\sigma = 1, 2, ..., 7$ frames, and differencing the smoothed series and use an off-the-shelf Röder peak picking algorithm \cite{DBLP:journals/jossw/MullerZ21} with standard parameters.
% We carry this out in the scenarios:
% % (ALL) use all annotated segment boundaries,
% % (CH) using only segment labels if they change,(FI) using only the first segment boundary of any segment labels.
% % %The two cases correspond to inclusions of different types of repetition: (ALL) considers all possible repetitions;
% % (CH) allows repetitions that are pairs of non-consecutive segments; and (FI) considers no repetitions.
% (FI) using only the first segment boundary of any segment labels,
% (CH) using only boundaries for which the corresponding segment labels change,
% (ALL) using all annotated segment boundaries.
% The three cases correspond to inclusions of different types of repetition.
% (FI) considers no repetitions and, as such, accounts for the boundaries of novel musical segments.
% (CH) allows repetitions that are pairs of non-consecutive segments, and accounts only for boundaries that change.
% (ALL) allows all type of repetition, and consider any segment boundary.  

\subsubsection{Pitch Surprisal and Uncertainty: IDyOM Consistency (SY)}
Although symbolic representations are largely limited to musical dimensions that can be discretized within specific (Western) notational traditions, they remain central to the study of musical expectation. This is because they capture core musical abstractions---such as discrete pitch and timing--- that are important to these traditions, have been thoroughly studied in music theory, and underpin a large body of perceptually validated findings.
We therefore test whether audio-based models yield MID estimates consistent with those from symbolic models, \textit{i.e.,} the SY task. Strong agreement would suggest that many findings on musical expectation established in the symbolic domain may also generalize to audio-based models. Following \cite{Bjare2025diffic}, we examine whether the MID measures IC and entropy estimated from audio correlate with those estimated from the perceptually validated symbolic pitch-expectancy model IDyOM \cite{idyom_conklin}.
This task is conducted on the SYN dataset.
For each context in the symbolic data, we extract the IDyOM MID measure and pair it with the same measure calculated with an audio-based model. To do so, we align the audio-based estimates to note events by locating the two audio frames pre/succeding the note onset and averaging their MID values.
We then evaluate their association using Spearman's rank correlation, since we expect the paired values to be related monotonically rather than linearly.

\subsubsection{Neural Encodings of MID measures (NE)}
% Testing whether estimated MID measures can predict behavioral and neural responses to natural music is crucial for linking computational models of music expectations to actual human perception. Indeed, demonstrating such predictive power provides evidence that these models capture fundamental predictive principles underlying the neural processing of musical structure \cite{di2020cortical}.
Testing whether estimated MID measures predict behavioral and neural responses to natural music is crucial for linking computational models of music expectation to human perception, providing evidence that these models capture predictive principles underlying neural processing of musical structure \cite{di2020cortical}.
To do so, NE relies on the multivariate Temporal Response Function (mTRF) framework \cite{mtrf} in which auditory processing hypotheses are framed as a model of how brain responses can be explained by continuous stimulus features, e.g., surprisal and uncertainty estimates, across multiple time lags, thereby also accounting for non-instantaneous interactions.
Such an approach is important, as we are probing neural processes that operate on the unfolding of musical structures over time in an ecologically valid scenario.

Similarly to \cite{Bjare2024audioic, Bjare2025perceptually}, NE uses Ridge regression for modeling EEG responses to VOC audio as a linear combination of two predictors: (i) surprisal and entropy (IC and $\mathcal{H}$
% as computed by each tested model
), (ii) the acoustic envelope of the waveform (E, Hilbert transform's magnitude).
Independent filters are estimated for each channel, participant, and model by solving a regularized linear regression with multiple time-lags \cite{mtrf}.
We chose a large $[-100, 700]$ ms window to safely account for all non-instantaneous interactions.

For each participant, model performance is evaluated using leave-one-out cross-validation across trials, and, to ensure normality, quantified as the z-transformed Pearson correlation $z$ between the predicted and observed EEG signals at each electrode.
% Significance of each feature's contribution is assessed as the difference in predictive power, $\Delta z$, between a full model and a reduced model that excludes the feature under test.
Feature significance is assessed as the difference in predictive power, $\Delta z$, between a full model and a reduced model excluding the tested feature.
Performance metrics are computed for each subject and channel, enabling model benchmarking at both the average and channel-wise levels while always assessing significance across subjects.
Additionally, we provide a function to select regions of interest (ROIs) based on the 10-20 system, enabling more targeted analyses.
% For instance, after inspecting spatial distributions with channel-wise performance plots on topographic maps (e.g., Figure~\ref{fig:topomaps}), one can look at global performance averaged over bilateral centrofrontal channels, known to produce reliable auditory responses \cite{Koelsch2012,DiLiberto2015}.
For instance, after inspecting spatial distributions via channel-wise performance topographic maps (e.g., Figure~\ref{fig:topomaps}), one can examine global performance averaged over bilateral centrofrontal channels, known for reliable auditory responses \cite{Koelsch2012,DiLiberto2015}.

Statistical analyses use one-tailed \textit{t}-tests or nonparametric Wilcoxon signed-rank tests, depending on the normality of the data, as assessed by the Shapiro-Wilk test. False discovery rate (FDR) is used to correct for multiple comparisons.
% Additionally, it is possible to compute a reference random distribution to test whether the observed statistic (e.g., the mean $r$ across subjects) deviates significantly from what would be expected under the null hypothesis of no association.
Optionally, a reference random distribution can be computed to test whether the observed statistic (e.g., mean $r$ across subjects) deviates significantly from the null hypothesis of no association.
The association is removed by randomly permuting stimulus-response pairs for the feature of interest $N_{perm}$ times.
% This procedure allows testing whether the improvement in the model's performance is really due to a significant relationship between a given feature and the responses, without introducing bias (e.g., by altering the model's complexity) \cite{di2020cortical}.
This tests whether the performance improvement reflects a significant relationship between a feature and the responses, without introducing bias (e.g., by altering model complexity) \cite{di2020cortical}.

In summary, NE produces a set of statistical and visual outputs for model benchmarking.
Predictive enhancement ($\Delta z$) is returned at three spatial resolutions: channel-wise, globally averaged, and, eventually, ROI-averaged, together with the statistical significance.
As a sanity check, the task reports channel-wise baseline $z$-scores to assess basic envelope tracking.
To support neurophysiological interpretation, NE generates a variety of visualizations, including bar plots of global performance, individual- and group-level scalp topographies masked by the significance of individual channels (significance threshold set at $p=0.05$), but also returns TRF weights for visual inspection.
Finally, MID measures can be used to predict other stimulus-aligned time series beyond EEG ---such as behavioral data---, thereby testing whether incorporating them significantly improves prediction accuracy.

\section{Experiments}

\label{sec:results}
In this section, we establish baseline performance on the benchmark suite to demonstrate how it can be used to evaluate and compare musical expectation models.
\\We extracted IC and entropy for the GIVT and DIF models on the datasets described in \Cref{sec:data} and performed all benchmarks as explained in \Cref{sec:tasks}.
DIF can estimate MID measures at different ``\textit{noise levels}'' $l$, which emphasize different acoustic or musical features.
For all our experiments involving DIF, we therefore report results for $l=0.0, 0.3, 0.5$, where $l=0.0$ retains full acoustic detail and higher levels emphasize more abstract features.
For DIF IC/entropy calculations, we use a hundred-step Euler ODE solver with four/one Skilling-Hutchinson trace estimator runs respectively (see \cite{Bjare2025diffic, Bjare2025perceptually} for details on the hyperparameters).
For calculating entropy in both models, we use 32 Monte Carlo samples.

\subsection{Reduction in MID measure with Repetition}
\begin{table}[!t]
\centering
\small
\setlength{\tabcolsep}{3pt}

\begin{tabular}{llccccccc}
\toprule
  & & & \multicolumn{2}{c}{Function} & \multicolumn{2}{c}{Uppercase} & \multicolumn{2}{c}{Lowercase} \\
 & & $l$ & ALL & FI & ALL & FI & ALL & FI \\ 
\cmidrule(r){4-5} 
\cmidrule(r){6-7} 
\cmidrule(r){8-9}
% \multirow{2}{*}{IC} & GIVT & -\textbf{.048} & -\textbf{.082} & -\textbf{.054} & -\textbf{.053} & -.030 & -.076 \\
%  & DIF & -.004 & -.074 & -.013 & -.048 & -.015 & -\textbf{.111} \\
% \cmidrule(r){3-4} 
% \cmidrule(r){5-6} 
% \cmidrule(r){7-8}
% \multirow{2}{*}{$\mathcal{H}$}& GIVT & -\textbf{.051} & -.077 & -\textbf{.061} & -.052 & -\textbf{.032} & -.068 \\
% & DIF & -.014 & -\textbf{.116} & -.025 & -\textbf{.093} & -.019 & -\textbf{.134} \\
\multirow{4}{*}{IC} & GIVT & - & -.048 & -.082 & -.054 & -.053 & -.030 & -.076 \\
\cmidrule(r){4-5} 
\cmidrule(r){6-7} 
\cmidrule(r){8-9}
 & \multirow{3}{*}{DIF} & .0 & -.019 & -.066 & -.031 & -.048 & -.019 & -.079 \\
 &  & .3 & -.021 & -.085 & -.033 & -.068 & -.021 & -.106 \\
 &  & .5 & -.004 & -.074 & -.013 & -.048 & -.015 & -.111 \\
\cmidrule(r){4-5} 
\cmidrule(r){6-7} 
\cmidrule(r){8-9}
\multirow{4}{*}{$\mathcal{H}$}& GIVT & -  & -.051 & -.077 & -.061 & -.052 & -.032 & -.068 \\
\cmidrule(r){4-5} 
\cmidrule(r){6-7} 
\cmidrule(r){8-9}
& \multirow{3}{*}{DIF} & .0 & -.011 & -.074 & -.028 & -.058 & -.018 & -.091 \\
&  & .3 & -.011 & -.100 & -.024 & -.078 & -.018 & -.120 \\
&  & .5 & -.014 & -.116 & -.025 & -.093 & -.019 & -.134 \\
\bottomrule
\end{tabular}
\caption{Baseline performances on RE showing the average decrease in MID measures with repetition. 
%Results are shown for all repetitions (ALL), and first contiguous repetitions (FI).
% $5\%$-unsignificant values in red.
}
\label{tab:repetition_results}
\end{table}
In \Cref{tab:repetition_results}, we report the results of the RE task.
All average differences are negative and significant on a $5\%$ level, and, as such, repetition reduces IC and entropy for all models at all annotation hierarchy levels, considering either all or first repetition occurrences (ALL and FI).
Comparing the repetition inclusion setups, DIF shows larger decreases for FI than for ALL, especially as the noise level $l$ increases.
This suggests that, for DIF, the reduction of these MID measures is concentrated more strongly on the first repetitions than on the later repetitions.
The same is generally true for GIVT, though to a lesser extent.
Across hierarchy levels, we find that the DIF model generally shows the highest reductions for the lowest level in the musical form hierarchy (Lowercase) for both repetition-inclusion setups (ALL or FI) and MID measures.
In fact, we find the highest overall reduction values among the models for IC and entropy at DIF lowercase FI.
When comparing IC and entropy with a fixed hierarchy level, model, and repetition-inclusion setup (ALL or FI), the largest decreases generally occur in entropy.

\subsection{Segment Boundary Alignment}
\begin{figure}[t]
    % \centering
    % \begin{minipage}[t]{0.33\linewidth}
        \centering
        \includegraphics[width=1.0\linewidth,trim=3.65cm 0.45cm 3.8cm 2.1cm,clip]{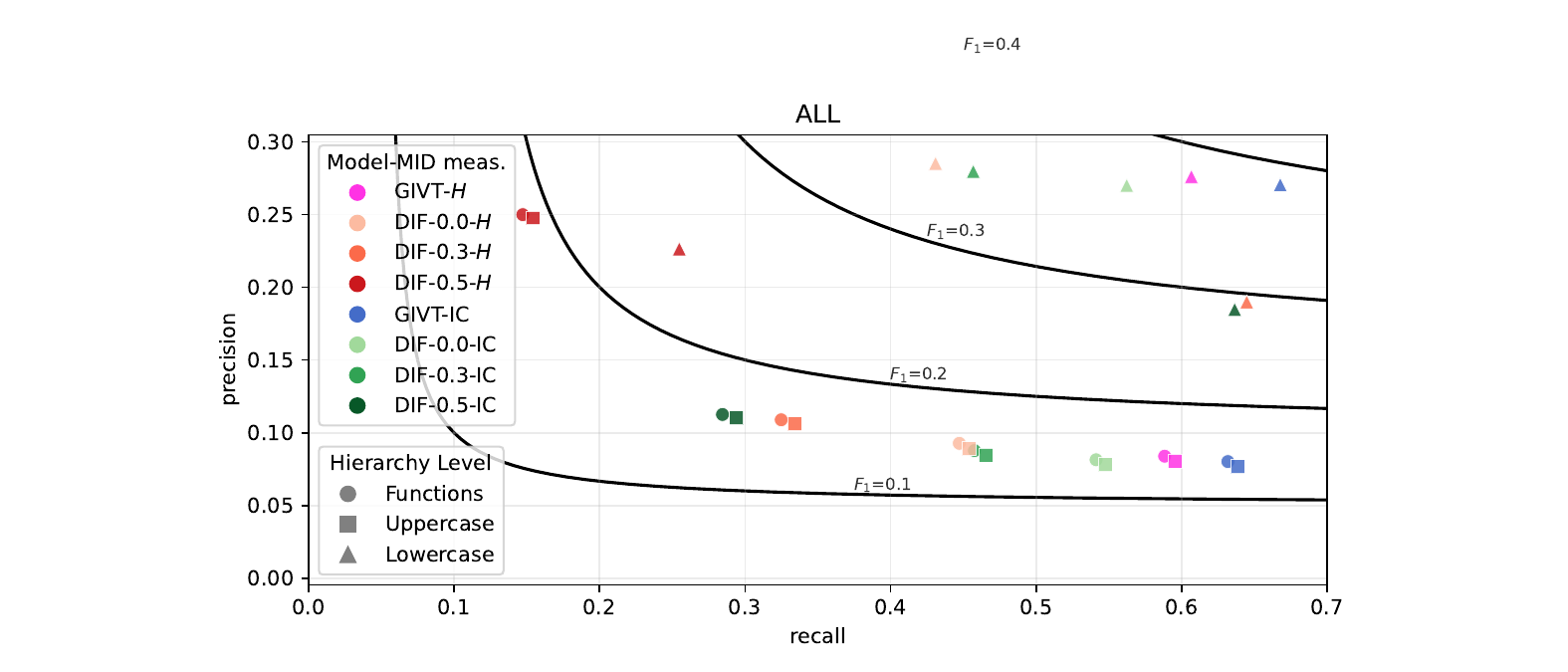}
    % \end{minipage}\hfill
    % \begin{minipage}[t]{0.33\linewidth}
    %     \centering
    %     \includegraphics[width=\linewidth,trim=0.5cm 0.25cm 0.5cm 0.25cm,clip]{figs/precision_recall_ch.pdf}
    % \end{minipage}\hfill
    % \begin{minipage}[t]{0.33\linewidth}
    %     \centering
    %     \includegraphics[width=\linewidth,trim=0.5cm 0.25cm 0.5cm 0.25cm,clip]{figs/precision_recall_fi.pdf}
    % \end{minipage}
    \caption{SB precision recall for the different hierarchy levels and models.}
    \label{fig:segboundary}
\end{figure}
In \Cref{fig:segboundary}, we report the results of SB.
A high recall indicates the extent to which the segment boundaries align with one peak in the MID measure.
Precision determines the degree to which peaks occur only around segment boundaries; low values indicate additional peaks, e.g., stemming from other musical events that are not associated with segment boundaries.
We interpret $F_1$ values as alignment scores, with $F_1=1$ corresponding to perfect alignment that exactly identifies all segment boundaries (and nothing else) by peaks in the MID measure.

The relatively higher alignment on Lowercase than on Functions and Uppercase, despite using optimal peak-picking parameters across all hierarchy levels, suggests that the peaks in the MID measures are more closely associated with lower-level segmentations than with higher-level ones.
Comparing the MID measures, IC generally shows higher alignment than entropy on Lowercase.
The fact that Uppercase and Functions are almost identical is expected, considering that segments in Functions are typically semantically relabeled versions of the segments in Uppercase and thus share segment boundaries.
Comparing the models, we find that GIVT has the highest alignment, and for the DIF model, higher noise levels decrease the alignment. 
Given the relatively low precision, many additional peaks are observed beyond those associated with segment boundaries.
% The fact that recall stays almost constant between the levels, together with the hierarchical segmentation in which Uppercase and Function boundaries are subsets of Lowercase boundaries, shows that many peaks detected near higher-level boundaries are already captured at the Lowercase level, while the lower precision at higher levels mainly reflects additional peaks aligned with lower-level boundaries that are not annotated at those coarser levels. 
% IC and entropy are found to perform similarily on the segment detection. 
% It is found that peaks in the GIVT model's entropy and IC align the most with segment boundaries.
% Thereafter, the DIF model that consider the full acoustic richness of the audio ($l=0.0$), with higher noise levels having the effect of increasing precision slighly, while lowering the recall a lot.  

\subsection{Pitch Surprisal and Uncertainty: IDyOM Consistency}
\begin{table}[!t]
\centering
\small
\setlength{\tabcolsep}{2.7pt}

\begin{tabular}{llccccccc}
\toprule
 & & $l$ & Piano & Guitar & Pad & Voice & Trumpet & Combined \\
\cmidrule(r){4-8}
\cmidrule(r){9-9}
% \cmidrule(r){9-9}
\multirow{4}{*}{IC} & GIVT & - & .439 & .428 & .432 & .373 & .457 & .390 \\
\cmidrule(r){4-8}
\cmidrule(r){9-9}
 & \multirow{3}{*}{DIF} & .0 & .487 & .550 & .550 & .584 & .561 & .489 \\
 &  & .3  & .520 & .563 & .578 & .617 & .596 & .523 \\
 &  & .5 & .521 & .534 & .556 & .512 & .571 & .490 \\

\cmidrule(r){4-8}
\cmidrule(r){9-9}
\multirow{4}{*}{$\mathcal{H}$} & GIVT & - & .439 & .497 & .442 & .530 & .539 & .463 \\
\cmidrule(r){4-8}
\cmidrule(r){9-9}
 & \multirow{3}{*}{DIF} & .0 & .529 & .577 & .538 & .604 & .569 & .522 \\
 &  & .3 & .532 & .562 & .537 & .607 & .581 & .520 \\
 &  & .5 & .506 & .515 & .507 & .564 & .554 & .493 \\
\bottomrule
\end{tabular}
\caption{Correlations between MID measures of audio models and those of IDyOM on SY.}
\label{tab:idyom_consistency_instruments}
\end{table}
In \Cref{tab:idyom_consistency_instruments}, we report the results for SY. To assess the models' sensitivities to timbre, we report correlations separately for each timbre and in the combined setting.
All correlations are significant on a $5\%$ level, and lie in the range 0.4-0.6.
Most variations in consistency are small. Entropy is slightly more consistent than IC.
Comparing the different timbres, we find that Piano generally produces the least consistent results.
The DIF model is more consistent than the GIVT model.
Similarly to the findings of \cite{Bjare2025perceptually}, the most consistent IC values of DIF are found for noise levels $l>0.0$. This, however, is not the case for entropy. 

\subsection{Neural Encodings of MID measures}

\begin{figure}[t]
    \centering

    \includegraphics[width=1\linewidth]{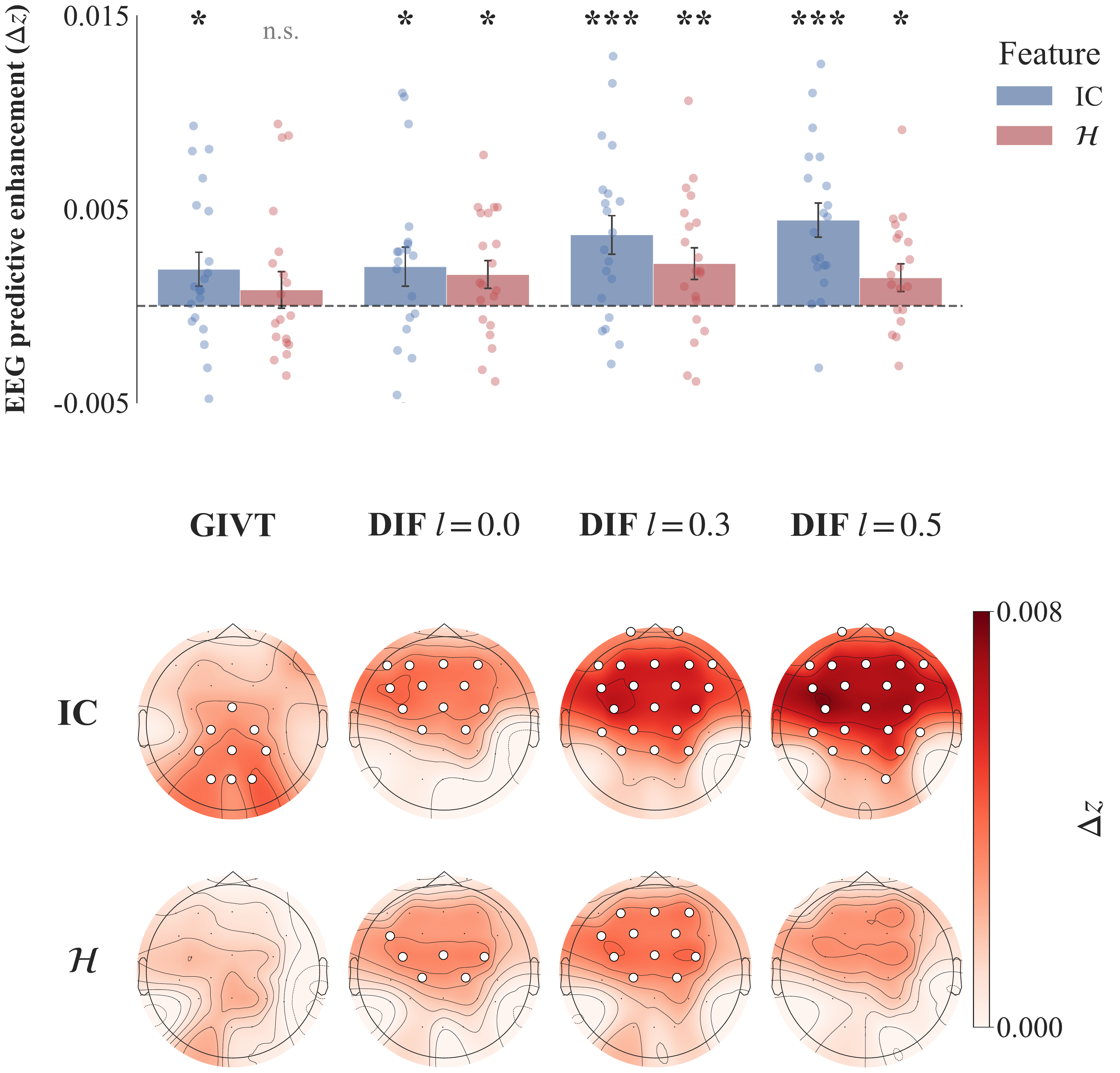}
    \caption {Bar plots report the z-score gain for IC and $\mathcal{H}$ (average across channels) over the baseline for each tested model averaged across all participants (mean ± SE). Dots represent individual participants. Topoplots report z-score gains across the scalp for the average participant. Only significant channels (FDR-corrected \textit{t}-test with significance threshold set at p < 0.05) are marked with white dots.\\}
    \label{fig:topomaps}
\end{figure}
Results of the NE task on the VOC dataset \cite{cantisani2023investigating} are illustrated in \Cref{fig:topomaps}. 
Different models are compared in terms of how much adding IC and $\mathcal{H}$ improves the prediction accuracy relative to a reduced model that includes only an acoustic control, in this case, the energy envelope of the waveform.
% As explained in section 3.4.4, the z-transformed correlation coefficients of the envelope-only baseline model predictions have been subtracted. 
% These alone were significant on all channels, with a jackknife t-test p-value smaller than 0.0001 everywhere and a mean $z_{base}=0.0356$. 
The improvement in predictive performance shows that brain responses track both IC and entropy for some of the models (\textit{t}-test or Wilcoxon based on normality of the data, GIVT IC: $p=0.0159$, $\mathcal{H}$: $p=0.4347$; DIF $l=0.0$ IC: $p=0.0266$, $\mathcal{H}$: $p=0.0138$; DIF $l=0.3$ IC: $p=0.0006$, $\mathcal{H}$: $p=0.0057$; DIF $l=0.5$: IC: $p<0.0001$, $\mathcal{H}$: $p=0.0168$), with IC generally being a more robust neural predictor than $\mathcal{H}$. Note that only a few survived the Bonferroni correction (adjusted $\alpha=0.05/8=0.00625$), specifically diffusion models DIF with higher level of noise (DIF $l=0.3$ and $l=0.5$). Note that the reduced model was significant per se (\textit{t}-test, $p<0.0001$), confirming the validity of the standard acoustic tracking baseline as a sanity check.
Most interestingly, when comparing different models, we found that higher diffusion noise levels for IC lead to significantly stronger encoding, similarly to the finding of \cite{Bjare2025perceptually}. %The predictive gain for IC peaked at the highest one (DIF $l=0.5$, $p < 0.001$), whereas contextual entropy was optimally tracked at a moderate level (DIF $l=0.3$).
% To directly compare models, we tested pairwise differences in subject-level global $\Delta z$ scores against zero. Normality was assessed via the Shapiro-Wilk test, followed by parametric paired 1-sample $t$-tests ($p_{Shapiro} > 0.05$) or non-parametric Wilcoxon signed-rank tests.
For IC, the model with the highest noise level (DIF $l=0.5$) outperformed all others, yelding greater neural tracking than GIVT ($\Delta z_{diff} = 0.0025, p < 0.001$), DIF $l=0.0$ ( $\Delta z_{diff} = 0.0024, p < 0.001$), and DIF $l=0.3$ ($\Delta z_{diff} = 0.0008, p = 0.047$), while entropy tracking peaked at a moderate noise level, with DIF $l=0.3$ outperforming DIF $l=0.5$ ($\Delta z_{diff} = 0.0007, p = 0.024$). However, only a few comparisons survived the Bonferroni correction (adjusted $\alpha=0.05/6=0.0083$).

\section{Discussion}
%This paper presents MEX as an initial step toward reproducible, cross-disciplinary work on musical expectation and computational models thereof. Rather than proposing a final account of musical expectation, we aim to provide a first set of tasks through which models can be evaluated against perceptually relevant criteria.
%In that sense, the main contribution is not only the individual benchmarks, but also the attempt to bring together perspectives that are often studied separately in MIR, music cognition, and neuroscience. We expect that some of the proposed tasks will prove more informative than others, and that both the tasks and their evaluation protocols will need to be refined as the field develops.
This paper presents MEX as an initial step toward reproducible, cross-disciplinary work on musical expectation.
In this initial version, the benchmarks test whether MID measures are sensitive to repetition learning, align with structural boundaries, agree across audio and symbolic domains, and predict neural responses.
Beyond the benchmarks themselves, MEX's main contribution is to bring together perspectives from MIR, music cognition, and neuroscience.
With this, we hope to further standardize perceptual validation of computational models of musical expectation and accelerate their development.

% As for the tasks, future work can include large-scale human ratings of musical expectation as ground truth. In particular, by leveraging the current audio-based design of the benchmark suite, comprehensive ratings for various real-world music can be integrated to rapidly test new models.
%
In addition to the proposed benchmarks, future work can focus on defining relevant behavioral tasks for which large-scale human data collection can be conducted.
For example, tasks such as the cloze probability paradigm, or continuous ratings of tension or surprisal during music listening, could be incorporated as online tasks providing a continuous data collection to further enrich the benchmark suite.
Furthermore, the current audio-based design of the benchmark suite, as opposed to a symbolic music design, facilitates the generalization of models to real-world music.

% In addition to benchmarking IC using computational and neural encoding metrics, it would be valuable to implement online behavioral tasks for large-scale data collection. For example, tasks such as the cloze probability paradigm, or continuous ratings of engagement or tension during music listening, could be incorporated to further validate and enrich the benchmark.
%So- standardize some behavioural tasks for an online experiment for a continuous collection of data.

%We expect the proposed tasks and evaluation protocols to be refined as the field develops.
A practical limitation of the current version is that the audio data of SAL, while commonly used in the community, is not strictly open access and has to be obtained through a separate, unstandardized procedure.
This remains an accessibility obstacle that we intend to address in subsequent versions, in part by adding new, diverse, open-access datasets (e.g., MTG-Jamendo) for the tasks.

In the current benchmark suite, tasks are evaluated in an unconditional zero-shot setting on music pieces without task-specific fine-tuning.
Future work could investigate other task specifications that enable perceptually supported ways to fine-tune, adapt, or condition models on task-specific context, thereby supporting the contextualization of multiple modalities where appropriate for particular musical expectation tasks.

Finally, as the number of tasks grows, it might become necessary to develop an ontology that groups the tasks according to core aspects of musical expectation, for example, in-context or corpus-based learning.
This would clarify which aspects a given model captures well.
We therefore see MEX as a starting point. We hope it provides a useful basis for more reproducible, collaborative work on musical expectation and its computational modeling.

% We emphasize that any other dataset following the same metadata structure could potentially be added to the benchmark in the future. The selection of specific datasets was a necessary step to standardize the tasks and ensure comparability across models and experiments. However, to fully assess the generalizability and robustness of the findings, it will be important to incorporate additional datasets in the future.

\clearpage
\section{Acknowledgments}
This work was supported by the European Research Council (ERC) under
the European Union’s Horizon 2020 research and innovation programme,
grant agreement 101019375 (“Whither Music?”).
% \section{AI Usage Statement}

% In line with the ISMIR 2026 AI Usage Policy, if you need to declare and explain the use of AI tools in the work described in the paper, you can do so by including a section titled ``AI Usage Statement" at the end of the paper. The AI Usage Statement section can be included both at submission time and in your camera-ready version. See the Call for Papers and Author Guidelines for more details. The AI Usage Statement section does \textit{not} count towards the page limit for scientific content. 

% \section{Ethics Statement}

% You may include an optional Ethics Statement section at the end of the paper to provide additional ethical considerations related to your paper. The Ethics Statement section can be included both at submission time and in your camera-ready version. See the Call for Papers and Submission Guidelines for details. The Ethics Statement section does \textit{not} count towards the page limit for scientific content.

% For BibTeX users:
\bibliography{ISMIRtemplate}
%TC:endignore
% For non BibTeX users:
%\begin{thebibliography}{citations}
% \bibitem{Author:17}
% E.~Author and B.~Authour, ``The title of the conference paper,'' in {\em Proc.
% of the Int. Society for Music Information Retrieval Conf.}, (Suzhou, China),
% pp.~111--117, 2017.
%
% \bibitem{Someone:10}
% A.~Someone, B.~Someone, and C.~Someone, ``The title of the journal paper,''
%  {\em Journal of New Music Research}, vol.~A, pp.~111--222, September 2010.
%
% \bibitem{Person:20}
% O.~Person, {\em Title of the Book}.
% \newblock Montr\'{e}al, Canada: McGill-Queen's University Press, 2021.
%
% \bibitem{Person:09}
% F.~Person and S.~Person, ``Title of a chapter this book,'' in {\em A Book
% Containing Delightful Chapters} (A.~G. Editor, ed.), pp.~58--102, Tokyo,
% Japan: The Publisher, 2009.
%
%\end{thebibliography}

\end{document}